\documentclass[twocolumn,prb,showpacs,superscriptaddress]{revtex4} 
\usepackage{color}
\usepackage{graphicx}
\usepackage{dcolumn}
\usepackage{bm}
\usepackage{amsmath}
\usepackage{nicefrac}

\RequirePackage{xspace}
\newcommand{\ie}{\textit{i.e.}\xspace}
\newcommand{\Ie}{\textit{I.e.}\xspace}
\newcommand{\eg}{\textit{e.g.}\xspace}

\newcommand{\YBCO}{\ensuremath{\mathrm{YBa_2Cu_3O_{7-\delta}}}\xspace}
\newcommand{\NCCO}{\ensuremath{\mathrm{Nd_{2-x}Ce_{x}CuO_{4-\delta}}}\xspace}

\newcommand{\opi}{\ensuremath{0\text{\,-\,}\pi}\xspace}

\newcommand{\av}[1]{\ensuremath{\left\langle #1 \right\rangle}}
\newcommand{\OPII}[1][4]{\ensuremath{#1\times(\opi \, \text{-})}}
\newcommand{\kreis}[1]{\unitlength1ex\begin{picture}(2.5,2.5)%
\put(0.75,0.75){\circle{2.5}}\put(0.75,0.75){\makebox(0,0){#1}}\end{picture}}

\begin{document}
\title{Magnetic field dependence of the critical current in \YBCO/Au/Nb ramp-zigzag Josephson junctions}
\author{S.~Scharinger}
\author{M.~Turad}
\author{A.~St\"{o}hr}
\affiliation{%
  Physikalisches Institut and Center for Collective Quantum Phenomena in LISA$^+$,
  Universit\"at T\"ubingen, Auf der Morgenstelle 14,
  D-72076, T\"ubingen, Germany
}

\author{V.~Leca}
\affiliation{%
  Physikalisches Institut and Center for Collective Quantum Phenomena in LISA$^+$,
  Universit\"at T\"ubingen, Auf der Morgenstelle 14,
  D-72076, T\"ubingen, Germany
}
\affiliation{%
  National Institute for Research and Development in Microtechnologies, 
  Molecular Nanotechnology Laboratory, 
  Erou Iancu Nicolae Str. 126A, RO-077190, Bucharest, Romania
}

\author{E.~Goldobin}
\affiliation{%
  Physikalisches Institut and Center for Collective Quantum Phenomena in LISA$^+$,
  Universit\"at T\"ubingen, Auf der Morgenstelle 14,
  D-72076, T\"ubingen, Germany
}

\author{R. G.~Mints}
\affiliation{The Raymond and Beverly Sackler School of Physics and Astronomy, Tel Aviv University, Tel Aviv 69978, Israel}

\author{D.~Koelle}
\author{R.~Kleiner}
\email{kleiner@uni-tuebingen.de}
\affiliation{%
  Physikalisches Institut and Center for Collective Quantum Phenomena in LISA$^+$,
  Universit\"at T\"ubingen, Auf der Morgenstelle 14,
  D-72076, T\"ubingen, Germany
}
\date{\today}

\begin{abstract}
We study the critical current $I_c$ dependence on applied magnetic field $H$ for multifacet \YBCO-Au-Nb ramp-type zigzag Josephson junctions. For many experiments one would like to apply a homogeneous field in the junction plane. However, even tiny misalignments can cause drastic deviations from homogeneity. We show this explicitly by measuring and analyzing $I_c$ vs. $H$ for an 8 facet junction, forming an array of \OPII \,-\,segments. The ramp angle is $\theta_r=8^\circ$. 
$H$ is applied under different angles $\theta$ relative to the substrate plane and different angles $\phi$ relative to the in-plane orientation of the zigzags. 
We find that a homogeneous flux distribution is only achieved for an angle  $\theta_h\approx 1^\circ - 2^\circ$ and that even a small misalignment $\sim$ 0.1$^\circ$ relative to $\theta_h$ can cause a substantial inhomogeneity of the flux density inside the junction, drastically altering its $I_c$ vs. $H$ interference pattern. We also show, that there is a dead angle $\theta^*_d$ relative to $\theta_h$ of similar magnitude, where the average flux density completely vanishes.\\
\end{abstract}


\maketitle
\subsection{Introduction}
\begin{figure}[tb]
\begin{center}
\includegraphics[width=0.85\columnwidth,clip]{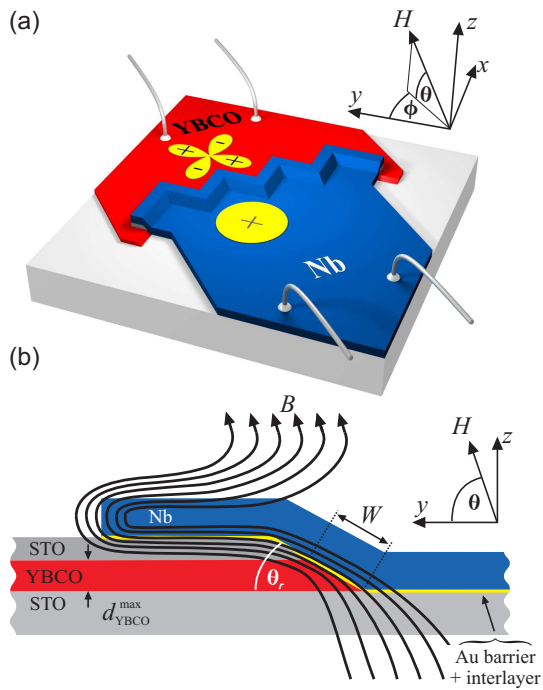}
\end{center}
\caption{(Color online). Sketch of ramp-zigzag junction: (a) whole junction with 8 facets, (b) cross section of one facet. Penetration and focusing effect of applied magnetic field along the ramp are indicated by black lines. 
Coordinates, and ramp angle $\theta_r$, 
as well as angles $\theta$ and $\phi$ of applied field $H$ are indicated.} 
\label{fig:sketch}
\end{figure}

Large (in a geometrical sense) Josephson junctions (JJs) are studied intensively since decades, \eg in the context of Josephson fluxon physics. Interesting and important physics is related to junctions where some regions obey the usual Josephson relation, while other regions incorporate an additional phase jump of $\pi$, which can be viewed as a negative critical supercurrent density $j_c<0$. These \opi junctions can be fabricated in different ways, \eg by connecting a d-wave superconductor such as \YBCO (YBCO)\cite{VanHarlingen95,Smilde02,Hilgenkamp03,Guerlich09} or \NCCO (NCCO)\cite{Guerlich09,Ariando05} to a conventional superconductor like Pb or Nb, or by using biepitaxial grain boundaries in YBCO\cite{Cedergren:2010:TTGB:0-pi-JJ:Semifluxon,Cedergren:2010:TTGB:0-pi-JJ,Stornaiuolo10, Stornaiuolo11, Longobardi12}, or superconductor-ferromagnet-superconductor (SFS) 
or superconductor-insulator-ferromagnet-superconductor (SIFS) junctions  \cite{Weides06a,Weides07a,Wild10}. 
In the YBCO-Nb or NCCO-Nb structures the Josephson junction is often of the ramp type and the barrier forms a zigzag line parallel to the crystallographic $a$ and $b$ axes \cite{Smilde02,Hilgenkamp03,Ariando05,Guerlich09}, see Fig.~\ref{fig:sketch}. This type of junction has been important for determining the symmetry of the order parameter of the cuprate superconductors \cite{VanHarlingen95,Tsuei00}. 

In addition, several properties make zigzag junctions very interesting for ongoing studies. In particular, half-integer vortices (semifluxons) can form spontaneously at the corners of the zigzag line and, thus, quasi-one-dimensional vortex crystals can be realized \cite{Hilgenkamp03,Susanto05}. Also, under special conditions the zigzag junctions can be used to create $\varphi_0$\cite{Buzdin08,Gumann09} or $\varphi$ junctions, with an arbitrary value of the ground state phase $\varphi$\cite{Buzdin03,Goldobin07a,Zazunov09,Goldobin11}. 
Such junctions, if long compared to the Josephson penetration depth $\lambda_J$, can carry mobile fractional vortices (splintered vortices) having many unusual properties \cite{Mints98,Mints01,Mints02,Moshe07b,Goldobin07a}. 
Apart from these research areas, aiming mostly at long junctions, ramp junctions are also interesting for superconducting electronics, \eg in the context of self-biased RSFQ circuits \cite{Ortlepp06} or in the context of superconducting quantum interference filters \cite{Oppenlaender03,Seifried05}. 
In all cases a good understanding of the implications of the zigzag ramp geometry is required.

Studies of Josephson junctions 
in many cases require the application of a magnetic field $H$ which, in theoretical studies, is usually considered to be oriented ``parallel'' to the junction plane, 
leading to a homogeneous flux density $\mu_0H$ in the absence of self fields generated by the Josephson currents. Most Josephson junctions have a simple geometry where the barrier layer and also the superconducting layers are oriented parallel to the substrate plane. Then, ``parallel'' simply means parallel to the substrate plane ($\theta = 0^\circ$, using the coordinates defined in Fig.~\ref{fig:sketch} (a)). For ramp junctions the junction plane (barrier layer) is tilted relative to the substrate plane by the ramp angle $\theta_r$ ($\theta_r = 8^\circ$ for the junction we study here), while the superconducting layers are partially parallel to the substrate plane and partially bent along the ramp, cf. Fig.~\ref{fig:sketch} (b). ``Parallel'' is thus not defined well. As it will be discussed in detail in Sec. C, there is an angle $\theta_h$, with $0<\theta_h<\theta_r$, where the applied field creates an almost homogeneous and uncompressed flux density in the junction. A magnetic field applied under this angle should be referred to as parallel.

Experimental studies on the zigzag junctions have usually been performed with the magnetic field applied perpendicular to the substrate plane \cite{Smilde02,Hilgenkamp03,Ariando05,Guerlich09}. One reason for this was the finding that, when aligning the 
field roughly parallel to the barrier layer, apart from a field scaling factor due to flux focusing, no essential difference to the perpendicular field orientation was observed. 
Theoretical interference patterns $I_c$ vs. $H$, calculated under the assumption that the applied field causes a homogeneous flux density $B$ in the junction plane
, and the experimental patterns agreed 
only qualitatively: the critical current $I_c$ was maximum when the flux per (\opi)\,-\,segment roughly equalled one flux quantum \cite{Smilde02,Hilgenkamp03, Ariando05,Guerlich09}. However, in almost any other respect experimental and theoretical $I_c$ vs. $H$ curves were not even similar. 

For junction geometries where the barrier layer is oriented either parallel or perpendicular to the substrate plane it was shown previously that for a field orientation perpendicular to the substrate plane ($\theta = 90^\circ$) the flux density in the junction barrier becomes inhomogeneous in the case of a homogeneous applied field\cite{Monaco08a,Monaco09, Moshe09, Scharinger10, deLuca11}. This leads, \eg, to a striking difference of $I_c$ vs. $H$ for SIFS multifacet \opi junctions (SIFS-MJJs), when measured in fields applied parallel and perpendicular to the substrate plane\cite{Guerlich10, Scharinger10}. Moreover, at least for junctions with a barrier layer oriented parallel to the substrate plane (for which $\theta_h=0$) and $H$ applied under an arbitrary angle $\theta$ 
there is a ``dead angle'' $\theta_d$ where the magnetic flux caused by the parallel ($\theta = 0^\circ$) and perpendicular ($\theta = 90^\circ$) components 
of $H$ cancel, leading to a critical current which almost does not modulate with $H$ \cite{Heinsohn01,Monaco09,Scharinger10}. The dead angle $\theta_d$ can be very close to zero, making proper junction alignment parallel to $H$ very difficult, if not impossible.

Obviously, ramp zigzag junctions are considerably more complex than conventional overlap junctions or SIFS-MJJs and need separate consideration. A systematic study under oblique fields seems necessary, having in mind that this type of junction is very useful for many future investigations. In the present study we have chosen a YBa$_2$Cu$_3$O$_7$-Au-Nb junction with 8 facets (\OPII \,-\,segments), each facet being 10\,$\mu$m long. We have investigated its $I_c(H)$ dependence as a function of $\theta$ and also the in-plane angle $\phi$, where  $\phi = 0^\circ$ corresponds to $H$ applied along the facets oriented in $y$ direction (cf. Fig.~\ref{fig:sketch}). 
Below we show that the field component perpendicular to $\theta_h$ leads to a periodically modulated flux density profile. The average value of the flux density caused by this component is enormously compressed, by a factor $\approx 100$ compared to $B=\mu_0 H$. 
As a consequence, the dead angle $\theta^*_d$, measured relative to $\theta_h$, 
is very small, $\theta^*_d \approx -0.38^\circ$ for $\phi = 45^\circ$ and  $\theta^*_d \approx -0.25^\circ$ for $\phi = 0^\circ$.
To achieve a more or less homogeneous flux density $B$, the field $H$ must be aligned better than some $0.1^\circ$ relative to $\theta_h$ and, to achieve 1\,$\Phi_0$ or more per (\opi)\,-\,segment, $\mu_0H$ values of more than 3\,mT are required. These conditions require quite dedicated experimental setups. Alternatively, realistic theoretical analyses should be based on the case of $\theta \approx 90^\circ$.

\subsection{Samples and Measurement Techniques}
%
The YBCO layer was grown by pulsed laser deposition (PLD) on a $[001]$-oriented SrTiO$_3$ (STO) single crystal substrate. The substrate temperature $T_s$ during the deposition of the 65\,nm thick YBCO thin films was $770^\circ$C and the oxygen pressure $P_{O_2}$ was 20\,Pa. A 60\,nm thick STO insulation layer on top of YBCO was also deposited by PLD at $T_s=760^\circ$C and $P_{0_2}=9\,$Pa. The targets were ablated using a KrF excimer laser at a repetition rate of 2\,Hz. After thin film deposition, the chamber was vented with oxygen up to 0.5\,bar and the sample was cooled down to room temperature, with an annealing step at $450^\circ$C for 30\,min. The zero-resistance transition temperatures $T_c$ of the YBCO films were between 88 and 90\,K. The YBCO/STO bilayer was patterned using optical lithography and Ar-ion milling under an angle of $30^\circ$ between surface normal and ion beam. To ensure a constant ramp angle for all junction orientations the sample was rotated about the axis normal to the surface  during milling. With these parameters a ramp angle of $\theta_r=8^\circ$ was obtained.
After removal of the photoresist the sample surface was cleaned in a soft Ar rf-plasma, in situ. 
Before the next deposition steps, an interlayer of 4 unit cells of YBCO was grown by PLD with the same deposition and annealing parameters as before to provide recrystallization \cite{Smilde02a}. 
According to Ref.~[\onlinecite{Smilde02a}], the thin YBCO interlayer is expected to become superconducting only on the YBCO ramp area, but not on the STO substrate and insulation layer.
%
With electron beam evaporation a 9\,nm Au barrier was deposited, followed by a sputtered Nb layer of thickness $d_{\text{Nb}}$ = 100\,nm. The plasma-cleaning and the last three depositions steps were done without breaking the vacuum. The Nb and Au layer were patterned by a final photo lithography and Ar-ion milling step. 
The resulting JJ is sketched in Fig.~\ref{fig:sketch}.
%


The samples were measured at $T$ = 4.2\,K in a magnetically and electrically shielded cryostat. The mounted sample typically had a misalignment of the applied magnetic field relative to the substrate plane $\theta_{\rm{off}}$ below 1$^\circ$. 
An external field $\mu_0H$ of up to 3.5\,mT could be applied and continuously rotated with respect to $\theta$ via two perpendicular coils operated in linear combination. 
The in-plane angle $\phi$ was varied by mounting the sample with proper orientation relative to the in-plane field axis, resulting in a misalignment error $\phi_{\rm{off}} \sim 2-3^\circ$.

\begin{figure}[tb]
\begin{center}
\includegraphics[width=0.9\columnwidth,clip]{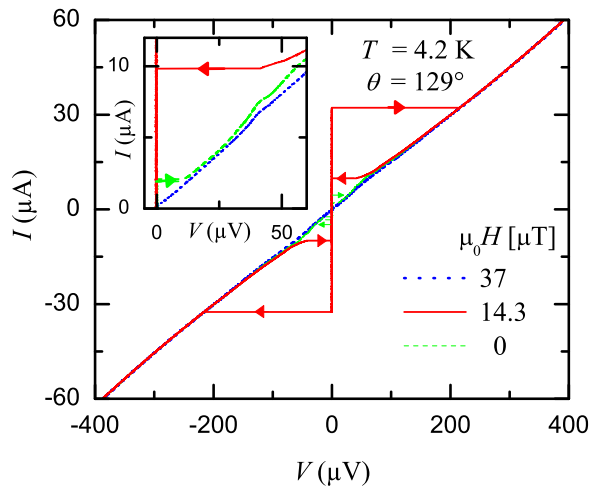}
\end{center}
\caption{(Color online). Current voltage characteristic of the 8-facet ramp zigzag JJ at $T$ = 4.2\,K for 3 different values of magnetic field, applied at $\theta=129^\circ$: 14.3\,$\mu$T (main $I_c$ maximum
), 0\,$\mu$T 
and 37\,$\mu$T ($I_c$ minimum
). Inset shows current voltage characteristic at expanded scales.}
\label{fig:iv}
\end{figure}

Below we discuss results from our most homogeneous sample. Fig.~\ref{fig:iv} shows current voltage characteristics, measured at three values of applied field. 
The current voltage characteristics  were hysteretic for critical currents $> 2\,\mu$A, with a junction resistance of $R\approx 6.6$\,$\Omega$. 
To measure $I_c$ the bias current $I$ was ramped up at fixed magnetic field until the junction switched to its resistive state. A voltage criterion $V_{\text{cr}}$ = 2\,$\mu$V was used to determine $I_c$, leading to a maximum over-estimate of $I_c$ by $V_{\text{cr}}/R \sim$ 0.3\,$\mu$A in the nonhysteretic regime. In the hysteretic regime $I_c$ is underestimated by some 0.1\,$\mu$A due to premature thermal activation.

\begin{figure}[tb]
\begin{center}
\includegraphics[width=0.8\columnwidth,clip]{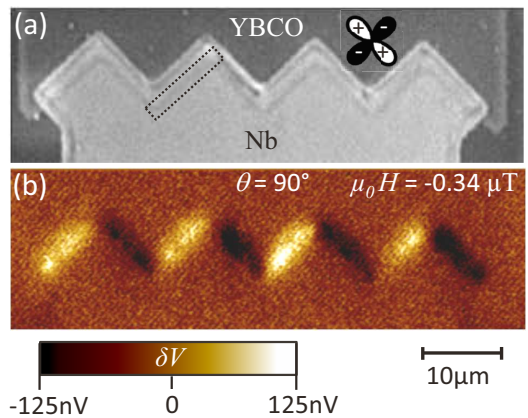}

\end{center}
\caption{(Color online) 
Images of the 8-facet YBCO-Nb zigzag JJ: (a) SEM surface image. The $d_{x^2-y^2}$-wave order parameter of the YBCO layer is indicated by the cloverleaf structure. White and black lobes are orientated along the crystallographic $a$ and $b$ axes and indicate the sign change of the order parameter. The dotted rectangle surrounds the ramp area of one facet.  (b) LTSEM $\delta V$ image taken at $T$ = 8.0\,K, $I$ = 2.3\,$\mu$A and $\mu_0H = -0.34$\,$\mu$T (central $I_c$ minimum at small offset field for $\theta = 90^\circ$).}
\label{fig:LTSEM}
\end{figure}
%

%
Fig.~\ref{fig:LTSEM} (a) shows a scanning electron microscopy (SEM) surface image of the YBCO-Nb zigzag junction. The dotted rectangle surrounds the ramp area of one facet, which is visible as a faint grey line. The top Nb electrode overlaps the YBCO ramp area by extra 3\,$\mu$m (idle region). 
The ramp areas of all facets form a zigzag line parallel to the $a$,$b$ axes of the YBCO film.

To investigate the homogeneity of the critical current densities of individual facets $j_c^i$,
we have imaged the current distribution of the entire zigzag junction at zero magnetic flux using low temperature scanning electron microscopy (LTSEM). 
Details of the method can be found in Ref. \onlinecite{Guerlich09}. 
In brief, the pulsed focused electron beam, which is scanned across the sample, causes local heating $\delta T < $ 1\,K on a lateral length scale of $\approx$ 1-3\,$\mu$m, which determines the spatial resolution of this imaging technique. 
The measured integral quantity is the voltage $V$ across the junction, which is biased slightly above $I_c$. $\delta T$ results in a local reduction of $|j_c(T)|$. The corresponding change of the overall $I_c$ of the zigzag junction causes a slight change of $\delta V (x,y)$, which depends on the beam position $(x,y)$ on the sample surface. For a bias current slightly above $I_c$ and $B$ = 0, $\delta V(x,y)\propto -j_c(x,y)$. 
This imaging technique requires non-hysteretic current voltage characteristics, which for our junction is not the case for $T$ = 4.2\,K, cf. Fig.~\ref{fig:iv}. Therefore, the $\delta V(x,y)$ image of the YBCO-Nb zigzag junction in Fig.~\ref{fig:LTSEM} (b) has been taken at $T$ = 8.0\,K. The junction has been biased at $I$ = 2.3\,$\mu$A. A field $\mu_0H = -0.34$\,$\mu$T was applied to compensate the residual field in the LTSEM setup. 
The $\delta V$ image clearly shows the alternating sign of supercurrent flow across neighboring facets. 
On the scale of the spatial resolution of our imaging technique neither defects nor $j_c$ asymmetries between facets are visible. Furthermore, the critical current densities of the facets seems to be quite homogeneous.


\subsection{General Considerations}

\subsubsection{Single-Facet Ramp Junction}
\begin{figure}[tb]
\begin{center}
\includegraphics[width=0.90\columnwidth,clip]{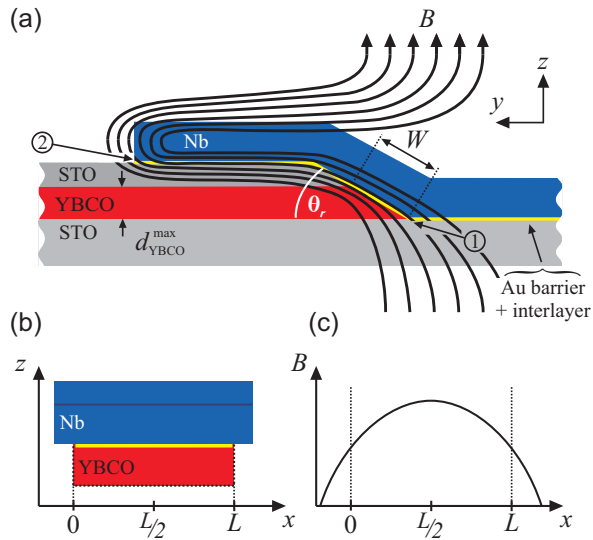}
\end{center}
\caption{(Color online). Sketch of a single facet JJ oriented along the $x$ axis, as considered in Sec. C:  
(a) Cross section parallel to the ($y,z$) plane. 
Penetration and focusing effect along the ramp for $H$ applied under $\theta=90^\circ$ are indicated by black lines. 
(b) Cross section in the ($x,z$) plane  at $y=(W/2)\cos\theta_r$ 
with (c) expected flux density profile $B(x)$ in the junction barrier. 
In the ($y,z$) plane $B$ is oriented along $\theta_r$.}
\label{fig:sketch2}
\end{figure}

Before we address $I_c(H)$ of zigzag JJs let us first theoretically consider a JJ with a single facet oriented parallel to the $x$ axis, 
(see Fig.~\ref{fig:sketch2}). 
%
The JJ area extends from $x=0$ 
to a length $L$ along the $x$ direction. 
The bottom YBCO electrode grows in thickness along $y$, reaching its maximum thickness $d^{\rm{max}}_{\rm{YBCO}}$ = 65\,nm at $y = d^{\text{max}}_{\rm{YBCO}}$/tan($\theta_r$) = 462\,nm, which is the projection of the junction width $W = d^{\rm{max}}_{\rm{YBCO}}$/sin($\theta_r$) = 467\,nm to the $y$ axis. 

Let us first estimate how the magnetic flux density distributes inside the junction when $H$ is applied perpendicular to the substrate plane ($\theta = 90^\circ$). The situation is sketched qualitatively in Fig.~\ref{fig:sketch2}.
We first note that $d^{\rm{max}}_{\rm{YBCO}}$ is well below the YBCO in-plane magnetic penetration depth $\lambda_{\rm{YBCO}} \sim 150$\,nm. 
Near the edge of the YBCO film located at $y = 0$  magnetic field lines can cross the YBCO film on the scale of the Pearl length \cite{Pearl64, Kirtley03a, Tafuri06} $\Lambda = 2 \lambda_{\rm{YBCO}}^2/d_{\rm{YBCO}} \gtrsim 700$\,nm.
This implies that over the whole junction area field lines can penetrate the YBCO film freely for any angle $\theta$.

By contrast, the thickness of the Nb film is comparable to the respective penetration depth ($\lambda_{\rm{Nb}}\sim$ 80\,nm) and approximately constant over the ramp area. Some field lines will cross the Nb film at its left edge (cf. Fig.~\ref{fig:sketch2} (a)) on a length scale of $\lambda_{\rm{Nb}}$.
However, due to the idle region this edge is far away from the junction area. Thus, no field line will cross the Nb film inside the junction area and the resulting flux density is essentially oriented parallel to the ramp ($\theta = \theta_r$).
Further, the screening currents preventing magnetic field lines from crossing the Nb layer lead to a strongly enhanced flux density in the junction barrier, cf. Fig.~\ref{fig:sketch2} (a). Along $x$ this effect is strongest at the center of the junction $(x = L/2)$. It disappears at the edge of the Nb film, cf. Fig.~\ref{fig:sketch2} (b,\,c). 
%
%

To account for these effects we may represent the resulting flux density profile inside the junction barrier, oriented along $\theta = \theta_r$, as
\begin{equation}
B_{90,f}(x)=\mu_0H f_{90,f} N_{90,f}\left[a_{90,f}+P_{90,f}(x)\right]
\label{Eq:B_90}
\end{equation}
%
where the field compression factor  $f_{90,f}\gg1$. The normalization constant $N_{90,f}$ is chosen such that the spatial average of $N_{90,f}(a_{90,f}+P_{90,f}(x))$ equals 1. 
The index `90' indicates perpendicular direction of the applied field ($\theta=90^\circ$) and the subscript `f' indicates the single facet scenario.
The spatially dependent field profile $P_{90,f}(x)>0$, to be specified later, is maximum in the center of the junction and is assumed to reach zero at its edges.
Without idle regions $B_{90,f}$ would drop to almost zero at the junction edges, \ie $a_{90,f}\,\approx\,0$. The effect of the idle regions is to make $B_{90,f}$ nonzero here. This is parametrized by $a_{90,f}$.

Using the spatial average $\av{P_{90,f}}$ of $P_{90,f}(x)$, $B_{90,f}$  can be regrouped as
\begin{equation}
B_{90,f}(x)=\mu_0H f_{90,f} \left[1+\widetilde{P}_{90,f}(x)\right]\,, 
\label{Eq:B_perp}
\end{equation}
where $\widetilde{P}_{90,f}(x) = N_{90,f}(P_{90,f}(x)-\av{P_{90,f}}) $ has vanishing average. 

The case of ``parallel'' fields requires some discussion. 
For theoretical considerations ``parallel'' should refer to an angle (relative to the substrate plane) where the flux density in the junction is homogeneous. 
We denote this direction as $\theta_h$. In the absence of idle regions a natural choice would be the ramp plane, \ie, 
$\theta_h=\theta_r$. Screening currents in the idle regions, however, will deform the flux density similar to the case of perpendicular fields, leading to an ambiguity of what ``parallel'' actually means. 

Let us consider a field $H$ applied at $\theta = 0^\circ$, \ie parallel to the substrate plane.
Screening currents in the vicinity of the ramp cause $B$ to be (almost) parallel to the Nb film, \ie tilted by $\theta_r$ within the area of the YBCO ramp. These screening currents also cause a slight field compression towards the center of the ramp. For the case of $\theta = 0^\circ$ we thus expect a flux density profile for the component along $\theta_r$ of the form 
\begin{equation}
B_{0,f}(x)=\mu_0Hf_{0,f} N_{0,f}\left[a_{0,f}+P_{0,f}(x)\right]
\label{eq:B_0}
\end{equation}
%
where the subscript `0' stands for $\theta = 0^\circ$. $f_{0,f} \gtrsim 1$ represents field compression. 
$N_{0,f}$ normalizes the field profile to 1. $P_{0,f}(x)$ has an absolute value which is maximum in the center of the facet and zero at its edges. The constant $a_{0,f}$ has been introduced to account for idle region effects. 

By decomposing $H$ into components perpendicular and parallel to the ramp one realizes that $P_{0,f}\leq0$ at least in the absence of idle regions. The screening currents in the idle regions reduce $|P_{0,f}(x)|$. 

Similar as $B_{90,f}$, $B_{0,f}(x)$ can be regrouped as
%
\begin{equation}
B_{0,f}(x)=\mu_0Hf_{0,f} \left[1+\widetilde{P}_{0,f}(x)\right]\,, 
\label{Eq:B_parallel}
\end{equation}
where $\widetilde{P}_{0,f}(x) = N_{0,f}(P_{0,f}(x)-\av{P_{0,f}}) $ has vanishing average. 

If $H$ is oriented within the $(y,z)$ plane at an arbitrary angle $\theta$ relative to the $y$ axis the total flux density in the junction is 
\begin{equation}
B_{\theta,f}(x)= B_{0,f}(x)\cos (\theta) + B_{90,f}(x)\sin (\theta) \,, 
\label{eq:B_theta}
\end{equation}
%
with $B_{0,f}(x)$ and $B_{90,f}(x)$ as defined in Eqs.~\eqref{Eq:B_parallel} and~\eqref{Eq:B_perp}. The factors $\cos(\theta)$ and $\sin(\theta)$ arise from a decomposition of $H$ into components parallel and perpendicular to the substrate plane. 

If the spatial dependences of $\widetilde{P}_{0,f}(x)$ and $\widetilde{P}_{90,f}(x)$ are similar, the ratio $p=|\widetilde{P}_{0,f}/\widetilde{P}_{90,f}|$ is about constant. Then, there is an angle 
$\theta_h$ = $-$arctan($f_{0,f}p/f_{90,f}$) where the flux density $B_{\theta,f}$ penetrating the junction is homogeneous along $x$ and given by $\mu_0H\cos \theta_h f_{0,f}(1-p/f_{90,f}) \approx  \mu_0H$. 

The angle $\theta_h$ might be referred to as ``field applied parallel to the ramp plane''\cite{Heinsohn01}. 
In the absence of idle regions we expect $\theta_h \approx \theta_r$. In their presence  $\theta_h$ is reduced. We did not perform an explicit calculation, but a guess is to consider a field line, which starts at the YBCO ramp edge at $y=0$ [point\, \kreis{1} in Fig.~\ref{fig:sketch2} (a)] and touches the edge of the overlapping Nb film [point\, \kreis{2} in Fig.~\ref{fig:sketch2} (a)], which, for $\phi=45^\circ$ (the relevant angle for the multifacet system), is at the projected length $\tilde{y}\approx$ 4\,$\mu$m and $z$ = 125\,nm. The corresponding angle is $\theta_h$ = 1.8$^\circ$.

A similar argument will hold for a multifacet system. In the data shown below we determined the misalignment angle relative to $\theta_h$ (``parallel'' alignment, homogeneous field) as $\theta^*_{\rm{off}} = 1.68^\circ$ at $\phi=45^\circ$ and as $\theta^*_{\rm{off}} = 0.85^\circ$ at $\phi=0^\circ$. 
We typically mount our samples  with an offset angle  $\theta_{\rm{off}} < 1^\circ$ relative to the substrate plane. Thus, we can put a limit $\theta_h \lesssim 2^\circ$ which is fully compatible with the above estimate of $\theta_h$ but rules out $\theta_h=\theta_r$. 


Finally, to account for the fact that $\theta_h$ cannot be precisely determined experimentally we introduce an angle $\theta^* = \theta-\theta_h$. By definition, the applied field is ``parallel'' for $\theta^*=0$ and ``perpendicular'' for $\theta^* \approx \theta = 90^{\circ}$. 

For the single facet case the flux density resulting from a field applied at an angle $\theta^*$ is:
\begin{equation}
B_{\theta^*,f}(x) = \mu_0H \cos(\theta^*) + \mu_0H \sin(\theta^*)g_{\perp,f} (x)\,,
\label{eq:B_theta_star}
\end{equation}
with 
$g_{\perp,f}(x) =  f_{\perp,f} [1+\widetilde{P}_{\perp,f}(x)]$. $f_{\perp,f}$ denotes the field compression. 
As in Eq.~\eqref{Eq:B_perp}, $\widetilde{P}_{\perp,f}(x) \approx \widetilde{P}_{90,f}(x)$ is a spatially varying function with zero average, having its maximum in the center of the junction plane at $x = L/2$. For the parallel field component ($\theta^* = 0^\circ)$ we have explicitly used that the flux density is homogeneous and essentially no flux compression occurs.


We have not yet made use of the fact that field lines can penetrate the YBCO film freely in the junction area. Thus, in $y$ direction the flux through the junction is not conserved but varies along $y$.
Alternatively, assuming a constant flux density along $y$, this can be rephrased in terms of an effective junction thickness $t_{\rm{eff}}$ that varies along the $y$ direction. 
Generally, $t_{\rm{eff}}$ can be found via \cite{Weihnacht69} 
\begin{equation}
t_{\rm{eff}} = t_{\rm{Au}} + \lambda_{\rm{YBCO}} \tanh \left(\frac{d_{\rm{YBCO}}}{2\lambda_{\rm{YBCO}}}\right) + \lambda_{\rm{Nb}}\tanh \left(\frac{d_{\rm{Nb}}}{2\lambda_{\rm{Nb}}}\right).
\label{eq:teff}
\end{equation}
Over the ramp area, the YBCO film thickness $d_{\rm{YBCO}}$ grows along $y$ from 0 to $d^{\rm{max}}_{\rm{YBCO}}$. Thus, $t_{\rm{eff}}$ increases from $\sim 53$\,nm to $\sim 85$\,nm over the ramp. For further calculations we use a linearized Ansatz 
\begin{equation}
t_{\rm{eff}}(y) = t_{\rm{eff0}}+\Delta t_{\rm{eff}}\frac{y}{W}\,,
\label{eq:teff_lin}
\end{equation}
with $-0.5<y/W<0.5$.
The average effective thickness $t_{\rm{eff0}}$ is given by Eq.~\eqref{eq:teff}, using $d_{\rm{YBCO}}\approx d^{\rm{max}}_{\rm{YBCO}}/2$.

Ignoring self-field effects caused by the Josephson current, the Josephson phase $\gamma$ is calculated via
%
\begin{equation}
\frac{d\gamma (x,y)}{dx}=\frac{2\pi}{\Phi_0}B(x)t_{\rm{eff}}(y).
\label{Eq:phase_parallel}
\end{equation}
The maximum supercurrent is obtained from 
%
\begin{equation}
I_c(H) = \max_{\gamma_0}\left \{ \iint dx\,dy\,j_c(x)\sin [\gamma(x,y)-\gamma_0]\right \}
\label{Eq:Ic_H}
\end{equation}
where integration has to be performed over the junction area. 
The dependence of $t_{\rm{eff}}$ on $y$ causes dephasing, which becomes severe when the flux \textit{difference} (along $y$) over the junction width becomes on the order of $\Phi_0/2$. 
In the above estimate $t_{\rm{eff}}$ differs by  $\Delta t_{\rm{eff}}/t_{\rm{eff0}} \approx 0.5 $ from its average value $t_{\rm{eff0}}$ ($\sim$69\,nm), and thus we expect the effect to become noticeable when the total flux through the junction becomes larger than about $\Phi_0$. However, the data discussed below for the 8 facet junction indicate $\Delta t_{\rm{eff}}/t_{\rm{eff0}} \approx 0.02$. Also, using $t_{\rm{eff0}}$ as a free parameter, our data indicate a value for $t_{\rm{eff0}}$ which is close to 85\,nm. \Ie only the ``thick'' part of the ramp seems to be relevant.
Eq.~\eqref{eq:teff} 
assumed isotropic superconductors and, thus, the YBCO anisotropy could lead to modifications. Whether or not this solves the discrepancy is unclear to us.

Let us summarize the central results of this section: We expect that in the junction plane the magnetic flux density along $x$ follows the shape given by Eq.~\eqref{eq:B_theta_star}. $B$ is homogeneous and equal to $\mu_0H$ if the field $H$ is applied under an angle $\theta_h$ ($\sim 1^\circ - 2^\circ$ for our geometry) relative to the substrate plane, which we denote as ``parallel field'' ($\theta^*=\theta-\theta_h=0^\circ$). Perpendicular components of $H$ lead to a compressed flux density which varies along $x$, having a maximum in the center of the junction at $x = L/2$. Due to the fact that flux lines can freely penetrate the YBCO film we expect that the total flux through the junction is not conserved but depends on $y$. This effect seems to be present in the zigzag junctions discussed below. However, its magnitude is at least an order of magnitude smaller than expected.

\begin{figure}[tb]
\begin{center}
\includegraphics[width=0.8\columnwidth,clip]{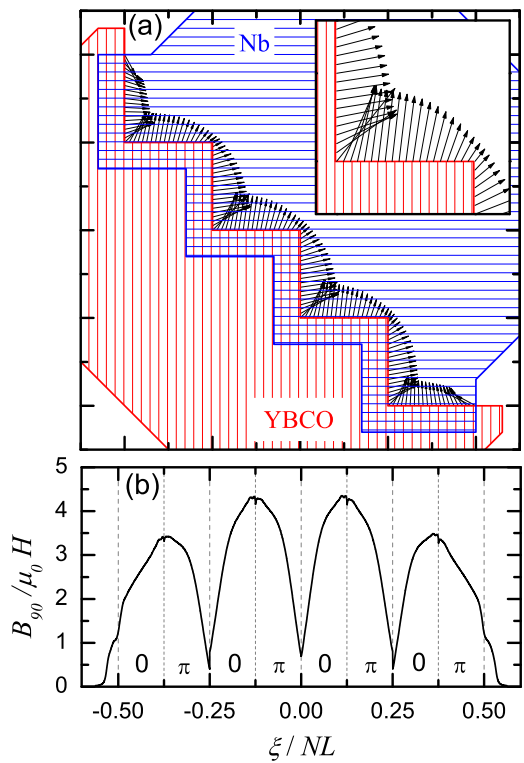}
\end{center}
\caption{(Color online). (a) Simulated in-plane magnetic field lines (black arrows) between two overlapping planar superconducting layers (YBCO, red vertical lines, Nb, blue horizontal lines) separated by a distance $d_L$ = 500\,nm. The magnetic field was applied perpendicular to the layers. Inset shows field lines at expanded scales. For clearness blue lines are omitted. (b) Flux density profile $B_{90}(\xi)$ of the calculated in-plane magnetic field $B$ projected onto the zigzag line. The coordinate $\xi$ runs along the zigzag line at the edge of the YBCO film.}
\label{fig:khapaev}
\end{figure}

\subsubsection{Multifacet Ramp Junction}
To get an idea of the field focused flux density profile inside the junction for $H$ applied at $\theta = 90^\circ$, we have simulated the flux density distribution around two overlapping planar superconducting layers located in the ($x,y$) plane being separated by a distance $d_L$ along $z$, using 3D-MSLI \cite{Khapaev03}, see hatched areas in Fig.~\ref{fig:khapaev} (a). The ramp was not included. The layer shape corresponded to the experimental situation and included $N=8$ facets of length $L$ = 10\,$\mu$m. 
We used $d_L$ = 500\,nm. 
Smaller values of $d_L$ led to convergence problems. The layer thicknesses were $d_{\rm{YBCO}}=65\,$nm and $d_{\rm{Nb}}=100\,$nm, respectively. 
Fig.~\ref{fig:khapaev} (a) shows the geometry projected onto the ($x,y$) plane and the field lines calculated in the plane parallel to the films and situated in the middle of the $d_L$ = 500\,nm gap between the films. The field was calculated along the edge of the YBCO film. 
$B$ is strongest at the inner edges of the YBCO layer and weakest at the outer edges. 
A flux density profile $B_{90}(\xi)$ $-$ we use the index `90' to indicate that $H$ has been applied perpendicular to the layers $-$ of the calculated in-plane magnetic field $B$ projected onto the zigzag line is shown in Fig.~\ref{fig:khapaev} (b).
$\xi$ is a curvilinear coordinate along the zigzag edge of YBCO. The projection is in units of $\mu_0H$ and the coordinate $\xi$ runs along the facets, $-0.5\leq \xi/NL\leq 0.5$. 
$B_{90}(\xi)/\mu_0H$ varies roughly sinusoidally, with one period per (\opi)\,-\,segment (by contrast, one might have expected one period per facet, cf. Fig.~\ref{fig:sketch2} (c)). $B_{90}(\xi)/\mu_0H$ reaches a maximum value of about 4.5, which is in fact much less than the actual field compression ($\sim 100$) found in experiment. It also turned out that the $B$ component \textit{perpendicular} to the layers is of the order of the in-plane component -- a feature which we expect to disappear in a more realistic scenario. Further, simulating $I_c(H)$ with the profile of  Fig.~\ref{fig:khapaev} (b) gave strong differences to the experimental interference pattern for flux values larger than $\Phi_0$ per (\opi)\,-\,segment. 

To simulate our zigzag junctions, as for the single facet case  we use angles $\theta^* =\theta - \theta_h$ and $\phi$ to describe the angles between the applied field and the junction. In this case ``perpendicular'' refers to $\theta^*=90^\circ$ and differs from $\theta=90^\circ$ by $\theta_h$. To account for this, below we use the index `$\perp$' to denote perpendicular direction. 
For the flux density generated by the perpendicular component of the applied field we follow the two-facet periodic shape obtained by the 3D-MSLI simulations but, in order to be more flexible, generalize it in the following way.
Each (\opi)\,-\,segment is described by a field profile $a_{\perp,f}+P_{\perp,f}(x)$,
with  $0 < x < 2L$. We used 
\begin{equation}
P_{\perp,f}(x)=\left(\sin\frac{\pi x}{2L}\right)^{\alpha_f}\,.
\label{eq:P_perp}
\end{equation}
%
The resulting periodic pattern along $\xi$, \ie the coordinate running along the zigzag line, is multiplied by an envelope function 
\begin{equation}
E(\xi)= a_{\perp,e}+P_{\perp,e}(\xi)
\label{eq:envelope}
\end{equation}
extending smoothly across the whole junction. We parametrized $P_{\perp,e}$ via 
\begin{equation}
P_{\perp ,e}=\left[1-\left(\frac{2\xi}{NL}\right)^2\right]^{\alpha_e}
\label{eq:P_envelope}
\end{equation}

The overall shape of $B_{\perp}(\xi)$ is given by
%
\begin{equation}
B_{\perp}(\xi)=\mu_0H f_{\perp} N_{\perp} E(\xi) \left[ a_{\perp,f}+P_{\perp,f}(\xi) \right] .
\label{Eq:B_perp_MJJ}
\end{equation} 
$N_{\perp}$ normalizes the spatial average of $B_{\perp}/\mu_0H f_{\perp}$ to 1, $f_{\perp}$ denotes field compression.
%

For arbitrary values of $\theta^*$ and $\phi$ the flux density $B$ through facets oriented parallel to $x$ is 
\begin{equation}
B_{\theta^*}(\xi)=B_{\perp}(\xi)\sin \theta^* + B_{\parallel} \cos \theta^* \cos \phi \,.
\label{eq:B_arbitrary_parallelx}
\end{equation}
The factors $\sin \theta^*$ and $\cos \theta^*\cos \phi$ arise from a projection of the applied field $H$.

For facets oriented parallel to $y$ one obtains 
\begin{equation}
B_{\theta^*}(\xi)=B_{\perp}(\xi)\sin \theta^* + B_{\parallel} \cos \theta^* \sin \phi \,,
\label{eq:B_arbitrary_parallely}
\end{equation}
with a homogeneous flux density $B_{\parallel}=\mu_0H$ along the zigzag line and $B_{\perp}(\xi)$ as defined in Eq.~\eqref{Eq:B_perp_MJJ}. 
Using this field profile (the actual parameters used are discussed in Sec. D, see also inset of Fig.~\ref{fig:int_gen} (a)) we solved Eqs.~\eqref{Eq:phase_parallel} and ~\eqref{Eq:Ic_H}, with $x$ replaced by $\xi$. The critical current density $j_c$ has been assumed to be constant in amplitude (homogeneous junction). It changes sign between adjacent facets.

Our model contains the 7 parameters $\alpha_f$, $\alpha_e$, $a_{\perp,e}$, $\Delta t_{\rm{eff}}$, $t_{\rm{eff0}}$, $f_{\perp}$ and $\theta_h$. Further, in experiment there is the offset angle $\theta_{\rm{off}}$ relative to the substrate plane.
As discussed in more detail in Sec. D, the shape of $I_c (H$) for ``perpendicular'' fields is not very sensitive to small errors in $\theta$ or $\theta^*$ when $\theta_h$ is small and thus, from this curve one obtains  $\alpha_f$, $\alpha_e$, $a_{\perp,e}$ and $\Delta t_{\rm{eff}}/t_{\rm{eff0}}$ with good accuracy. 
From the field modulation period in perpendicular field one further obtains the product $f_{\perp} t_{\rm{eff0}}$. A priori, $t_{\rm{eff0}}$ is not known precisely.  From the angle dependent $I_c(H)$ measurements one can identify the dead angle $\theta_d^*$ relative to $\theta_h$ with high accuracy. However, $\theta^* = (\theta - \theta_h) = 0^\circ$ is much harder to identify if, as in our case, $I_c(H)$ is measurable only over a small number of modulation periods. $\theta_d^*$ depends on $f_{\perp}$.  Thus, $f_{\perp}$  needs to be determined by comparing a set of $I_c(H)$ curves near parallel orientation, finally also allowing to determine $t_{\rm{eff0}}$ and the offset angle $\theta_{\rm{off}}^*$ relative to $\theta_h$. $\theta_h$ itself, however, is hard to determine and has an uncertainty which is on the order of the offset angle $\theta_{\rm{off}}$.

\subsection{Results}

\begin{figure}[tb]
\includegraphics[width=1.0\columnwidth,clip]{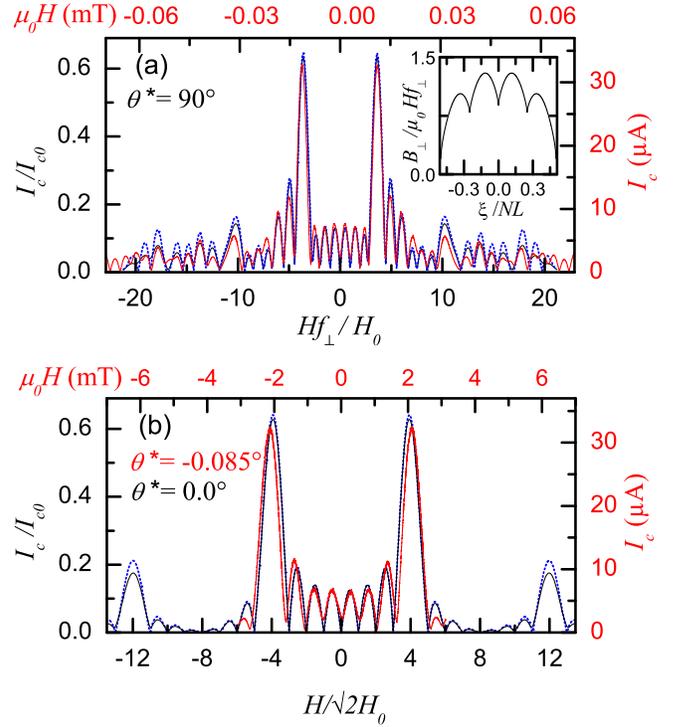}
\caption{(Color online).
Measured (red) and calculated (black, blue) interference patterns for (a) $\theta^* = 90^\circ$ and (b) $\phi = 45^\circ$ and $\theta^* = 0^\circ$ (calculation), $\theta^* = -0.085^\circ$ (measurement; offset angle $\theta^*_{\rm{off}} = 1.684^\circ$ subtracted).
$\theta^*=\theta - \theta_h=0$ corresponds to ``parallel'' field orientation, as defined in the text. 
Model parameters are $\alpha_e$ = 0.5, $a_{\perp,e}$ = 0.3, $\alpha_f$ = 0.7, $a_{\perp,f}$ = 2.0, $f_{\perp}$ = 100, cf. Eqs.~\eqref{eq:P_perp},~\eqref{eq:envelope},~\eqref{eq:P_envelope} and~\eqref{Eq:B_perp_MJJ}. For the black curve $\Delta t_{\rm{eff}}/t_{\rm{eff0}}$ = 0.02, for the dotted (blue) line $\Delta t_{\rm{eff}}$ = 0.
$H_0/f_{\perp}=NLt_{\rm{eff0}}/\Phi_0$. Critical current in the calculated plots is normalized to $I_{c0} = |j_c|A_J$, where $j_c$ is the critical current density and $A_J$ is the junction area. Inset in (a) shows flux density profile $B_{\perp}$ normalized to $\mu_0Hf_{\perp}$.
}
 \label{fig:int_gen}
\end{figure}
\par

Fig.~\ref{fig:int_gen} (a) compares measured and calculated interference patterns for $\theta^* = 90^\circ$. 
The horizontal scales of the calculated patterns are in units of $H_0/f_{\perp}$,  where $H_0 \equiv f_{\perp} NLt_{\rm{eff0}}/\Phi_0$, with the field compression factor for perpendicular field components $f_{\perp}$, the average effective junction thickness $t_{\rm{eff0}}$, the facet length $L$ and the facet number $N$.
The solid (black) line is for a relative variation of the effective junction thickness $\Delta t_{\rm{eff}}/t_{\rm{eff0}}$ = 0.02, the dotted (blue) line is for $\Delta t_{\rm{eff}}$ = 0.
The agreement between measured and calculated (for $\Delta t_{\rm{eff}}/t_{\rm{eff0}}$ = 0.02) interference patterns is reasonable, although differences occur for normalized fields $Hf_{\perp}/H_0\gtrsim5$. 
For $\Delta t_{\rm{eff}}/t_{\rm{eff0}} > $ 0.02, at high fields the $I_c$ maxima are suppressed in comparison with the $\Delta t_{\rm{eff}}/t_{\rm{eff0}}$ = 0.02 case, while the $I_c$ minima are significantly lifted from zero, both in contrast to the measurements. Thus the margin on $\Delta t_{\rm{eff}}/t_{\rm{eff0}}$ is relatively narrow.
For the (\opi)\,-\,segments we used in Eq.~\eqref{eq:P_perp} $\alpha_f$ = 0.7 and $a_{\perp,f}$ = 2.0. The power $\alpha_f$ was suggested by the 3D-MSLI simulations, cf. Sec. C. The calculated interference patterns, however did not depend strongly on this parameter. By contrast, the quite large value of $a_{\perp,f}$ was necessary to achieve a reasonable agreement with experimental data. As a result, the periodic modulations of $B_{\perp}(\xi)$ are much shallower than suggested by the  3D-MSLI simulations. 
For the envelope function (Eqs.~\eqref{eq:envelope} and~\eqref{eq:P_envelope}) we used $\alpha_e$ = 0.5 and $a_{\perp,e}$ = 0.3. Normalization resulted in $N_{\perp}$ = 0.34.
The corresponding field profile is shown in the inset of Fig.~\ref{fig:int_gen} (a).
Finally, by matching the abscissas of the theoretical and measured interference patterns we find $t_{\rm{eff0}} f_{\perp}$ = 8.5\,$\mu$m. A thickness $t_{\rm{eff0}}$ = 69\,nm calculated from geometry (film thicknesses) corresponds to $f_{\perp}$ = 120. In fact, for the $I_c(H)$ data at different values of $\theta^*$ and $\phi$ this value turned out to be somewhat too large. Best agreement was obtained for 
$f_{\perp} \sim 100$, corresponding to $ t_{\rm{eff0}} \sim $ 85\,nm. 
%

We mention here that a periodic modulation of $B_{\perp}(\xi$) with one period per facet rather than one period per (\opi)\,-\,segment also gave satisfactory agreement between calculated and measured interference patterns. Thus, from $I_c(H)$ we cannot unambiguously distinguish these scenarios.

Fig.~\ref{fig:int_gen} (b) shows measured and calculated interference patterns near parallel orientation 
for $\phi=45^\circ$.
For the calculation ($\theta^* = 0^\circ$) we have assumed that the magnetic flux density along the zigzag line of the multifacet junction $B_{\parallel}(x)$ is homogeneous and not compressed, \ie $B_{\parallel}(x)=\mu_0H$.
The field scale is given in units of $H_0 \sqrt{2}$. The factor $\sqrt{2}$ has been included to account for the fact that only a field $\mu_0H/\sqrt{2}$ is applied per facet. The main $I_c$ maximum in this normalization roughly appears at $N/2$. The dotted (blue) interference pattern is for $\Delta t_{\rm{eff}} =0$ to emphasize the ``ideal'' case. $\Delta t_{\rm{eff}}/t_{\rm{eff0}} =0.02$ (black line) yields only minor deviations in the field range shown.
The experimental curve is at an angle $\theta^*=-0.085^\circ$. No curve has been measured at $\theta^*=0.0^\circ$ an angle, which is defined only after data analysis. 
Note that, in principle, the experimental and theoretical field scales are linked for a fixed value of $f_{\perp}$. However, in order to compare the \textit{shape} of $I_c(H)$, in Fig.~\ref{fig:int_gen} (b) we have additionally compressed the experimental field scale by a factor of 1.2 to account for the nonzero value of $\theta^*$. 
Further note that the experimental curve only covers a limited field range. This is the limit set by our experimental system. 
Comparing the curves at $\theta^*=90^\circ$ and $\theta^*=0^\circ$ one notices that for $\theta^*=0^\circ$ the amplitudes of the secondary $I_c$ maxima situated between the two main $I_c$ maxima follow a U-shaped dependence while for $\theta^*=90^\circ$ their amplitudes are about constant. Also, for $\theta^*=0^\circ$ and absolute values of fields higher than the main $I_c$ maxima, the $I_c$ maxima are strongly reduced compared to $\theta^*=90^\circ$. These are the main differences in $I_c(H)$ at $\theta^*=0^\circ$ and $\theta^*=90^\circ$. Thus, the overall difference of the interference pattern at $\theta^*=90^\circ$ compared to the ideal case is much less striking than in previous publications \cite{Smilde02,Hilgenkamp03,Ariando05,Guerlich09}, presumably pointing to a much higher inhomogeneity of the critical current density in previous generations of ramp-zigzag \opi JJs.


\begin{figure}[tb]
\includegraphics[width=\columnwidth,clip]{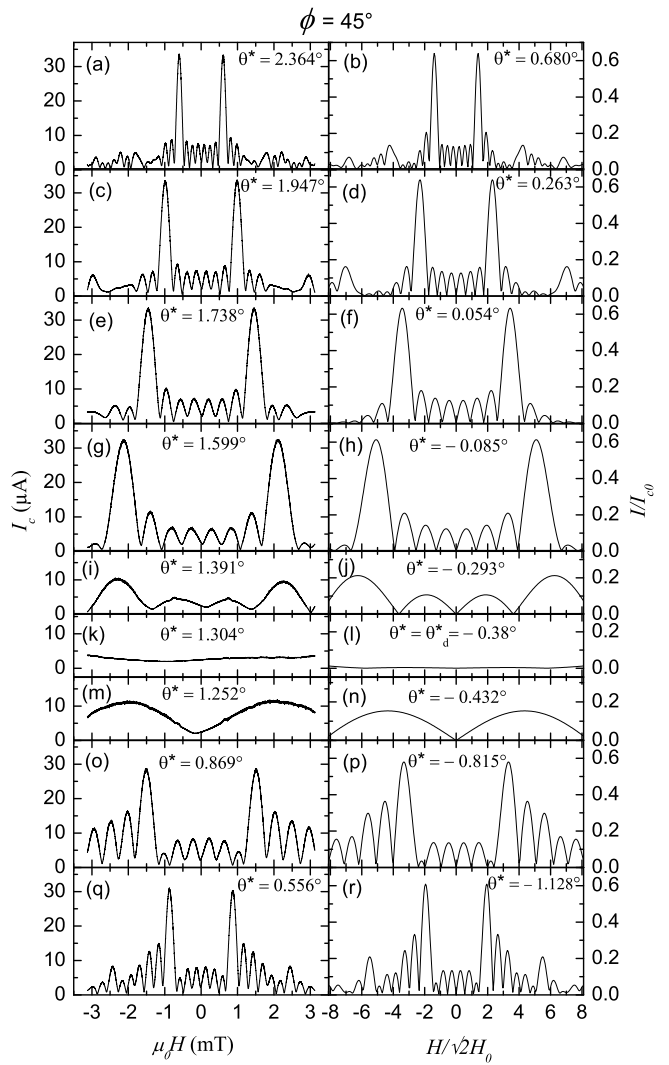}
\caption{Experimental (left) and calculated (right) interference patterns for $\phi = 45^\circ$ and $-1.128^\circ < \theta^* < 0.68^\circ$. $\theta^*=\theta - \theta_h=0$ corresponds to ``parallel'' field orientation, as defined in the text. Offset angle $\theta^*_{\rm{off}} = 1.684^\circ$ relative to $\theta^*=0$ is not subtracted in the experimental graphs. 
$H_0/f_{\perp} =NLt_{\rm{eff0}}/\Phi_0$. Critical current in the calculated plots is normalized to $I_{c0} =|j_c|A_J$, where $j_c$ is the critical current density and $A_J$ is the junction area.
Model parameters are $\alpha_e$ = 0.5, $a_{\perp,e}$ = 0.3, $\alpha_f$ = 0.7, $a_{\perp,f}$ = 2.0, $f_{\perp}$ = 100 and $\Delta t_{\rm{eff}}/t_{\rm{eff0}}$ = 0.02, cf. Eqs.~\eqref{eq:P_perp},~\eqref{eq:envelope},~\eqref{eq:P_envelope} and~\eqref{Eq:B_perp_MJJ}.
}
 \label{fig:int_45}
\end{figure}
\par

Fig.~\ref{fig:int_45} compares measured and calculated interference patterns for $\phi = 45^\circ$ and $-1.128^\circ < \theta^* < 0.68^\circ$, where the angles quoted here and also below refer to the calculated patterns. In the experimental curves the offset angle $\theta^*_{\rm{off}} = 1.684^\circ$ relative to $\theta^*=0$ is not subtracted. The agreement is fair for all angles shown. The only additional parameter required to fit the whole series of curves was $\theta^*_{\rm{off}}$. It was determined by comparing the calculated interference pattern at the dead angle, cf. Fig.~\ref{fig:int_45} (l), to the measured one, cf. Fig.~\ref{fig:int_45} (k).
For arbitrary $\phi$, the dead angle (relative to $\theta^*=0$) is given by 
\begin{equation} 
\theta^*_d \approx -\arctan \left(\frac{\cos \phi + \sin \phi}{2f_{\perp}}\right)\,, 
\label{eq:theta_d}
\end{equation}
with $\theta^*_d \approx -0.4^\circ$ for $\phi = 45^\circ$. Near $\theta^*_d$, variations in $\theta^*$ on the order of $0.01^\circ$ already cause significant changes in $I_c$ vs. $H$ so that, once $f_{\perp}$ is fixed, $\theta^*_{\rm{off}}$ can be determined very precisely. 
Note that the experimental $I_c$ vs. $H$ pattern of Fig.~\ref{fig:int_45} (k) slightly modulates around 2.5\,$\mu$A, while the calculated pattern in Fig.~\ref{fig:int_45} (l) is almost at zero current. This is presumably caused by residual fields in the cryostat, causing also the shift in the $I_c$ minimum relative to $H$ = 0 in Fig.~\ref{fig:int_45} (m). 
Also, the $\theta^*$ dependence of the interference patterns strongly depends on $f_{\perp}$. This parameter cannot be altered by more than some 5\,$\%$ from 100 without substantial degradation  of the fit qualitiy.  

\begin{figure}[tb]
\includegraphics[width=\columnwidth,clip]{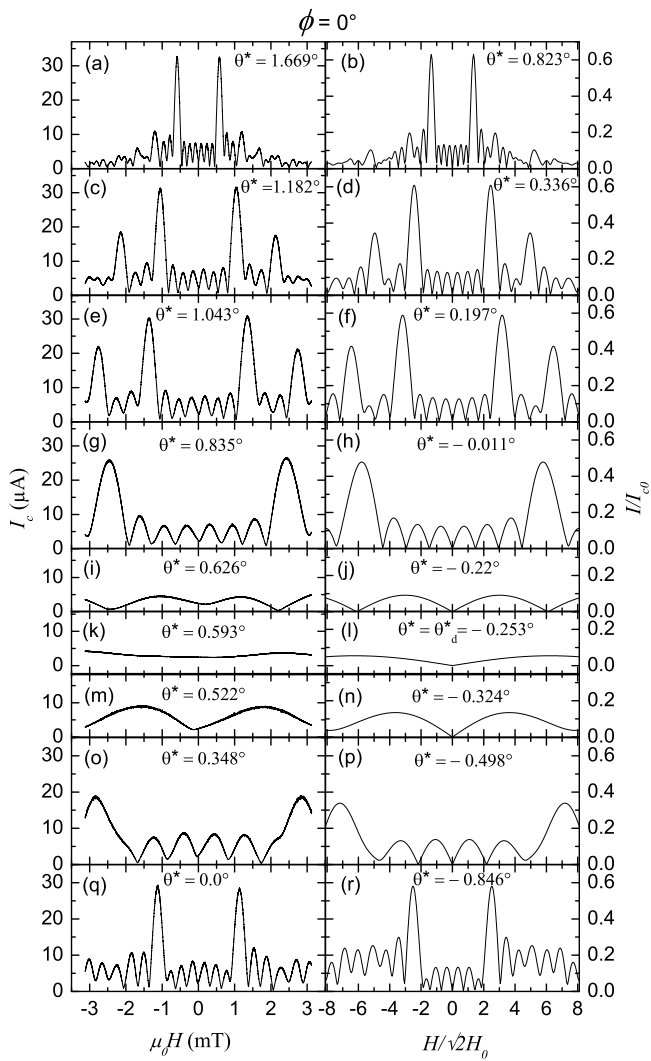}
\caption{Experimental (left) and calculated (right) interference patterns for $\phi = 0^\circ$ and $-0.846^\circ < \theta^* < 0.823^\circ$. $\theta^*=\theta - \theta_h=0$ corresponds to ``parallel'' field orientation, as defined in the text. Offset angle $\theta^*_{\rm{off}} = 0.846^\circ$ relative to $\theta^*=0$ is not subtracted in the experimental graphs. 
$H_0/f_{\perp} =NLt_{\rm{eff0}}/\Phi_0$. Critical current in the calculated plots is normalized to $I_{c0} =|j_c|A_J$, where $j_c$ is the critical current density and $A_J$ is the junction area.
Model parameters are $\alpha_e$ = 0.5, $a_{\perp,e}$ = 0.3, $\alpha_f$ = 0.7, $a_{\perp,f}$ = 2.0, $f_{\perp}$ = 100 and $\Delta t_{\rm{eff}}/t_{\rm{eff0}}$ = 0.02, cf. Eqs.~\eqref{eq:P_perp},~\eqref{eq:envelope},~\eqref{eq:P_envelope} and~\eqref{Eq:B_perp_MJJ}.
}
 \label{fig:int_0}
\end{figure}
\par

To further confirm that the profile $B(\xi)$ -- plus the assumption of homogeneous junction parameters -- describes the situation well we have also varied $\phi$.  Fig.~\ref{fig:int_0} compares data and calculations for $\phi = 0^\circ$ and various values of $\theta^*$ around $\theta^*=0^\circ$. For the case of $\phi=\theta^*=0^\circ$ calculations yield critical current main maxima $I_c = 0.5 I_{c0}$ whenever the flux through the facets oriented along $x$ equals a multiple of $\Phi_0$. Then the critical current of these facets cancel and $I_c$ is given by the sum of the critical currents of the facets oriented along $y$. Curves of Fig.~\ref{fig:int_0} (a)--(f) show the appearance of the first and second main $I_c$ maximum, which grows in amplitude for $\theta^* \rightarrow 0^\circ$. Although the maximum field provided by our setup was not sufficient to resolve the main maxima for $\theta^* = 0^\circ$ the evolution of the first and second $I_c$ main peaks are clearly visible for $\theta^* < 0.336^\circ$, with a good agreement between experimental and theoretical patterns. By lowering $\theta^*$ one again runs through a dead angle ($\theta^*_d = - 0.253^\circ$) and obtains somewhat strangely looking interference patterns for $-1^\circ < \theta^* < \theta^*_d$. 

\begin{figure}[tb]
\includegraphics[width=0.9\columnwidth,clip]{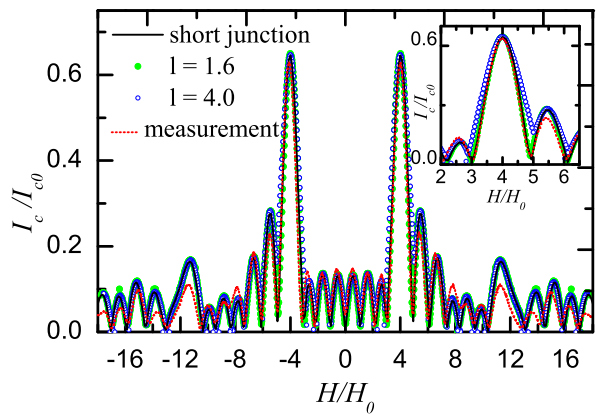}
\caption{(Color online). $I_c$ vs. $H$ ($\theta^* = 90^\circ$) patterns calculated with the short junction model for $\Delta t_{\rm{eff}}=0$ (solid  line) and simulated for different lengths $l$ (circles) using StkJJ\cite{StkJJ} in comparison with measurement (dotted line). Other model parameters are the same as in Figs.~\ref{fig:int_gen},~\ref{fig:int_45} and~\ref{fig:int_0}. Deviations between the calculated and the simulated (for $l=4$) pattern are visible in the zoomed inset.}
\label{fig:short-vs-long}
\end{figure}

Finally, we address the effect of self fields generated by the Josephson currents. Such effects become prominent when the junction is 2--3 times longer than the Josephson length $\lambda_J$. To estimate $\lambda_J$ we first ignore idle regions and assume that the supercurrent flows homogeneously across the junction area $A_J= NLW \approx 37$\,$\mu$m$^2$. We then find $j_{c0}=I_{c0}/A_J \approx $ 140\,A/cm$^2$. 
Using $\lambda_{J0} = [\Phi_0/2\pi\mu_0j_cd_{\rm{eff}}]^{0.5}$, with the effective magnetic junction thickness\cite{Weihnacht69}
\begin{eqnarray}
\nonumber d_{\rm{eff}} &=& t_{\rm{Au}} + \lambda_{\rm{YBCO}} \coth\left(\frac{d_{\rm{YBCO}}}{\lambda_{\rm{YBCO}}}\right)+ \lambda_{\rm{Nb}}\coth \left(\frac{d_{\rm{Nb}}}{\lambda_{\rm{Nb}}}\right) \\ & \approx & 470\,nm\,  
\label{eq:deff}
\end{eqnarray}
we find $\lambda_{J0} \approx$ 20\,$\mu$m and the normalized junction length $l = NL/\lambda_J \approx$ 4.
The idle region effect \cite{Monaco95, Maggi97} leads to an increased Josephson length, $\lambda_{J\rm{,i}} = \delta \cdot \lambda_{J0}$, with  $\delta=(1+(d_{\rm{eff}}/d_{\rm{eff,i}})(W_i/W))^{0.5}$. $W_i$ and $d_{\rm{eff,i}}$, respectively, are the width and effective magnetic thickness of the idle region. 
With $d_{\rm{eff,i}}$ = 535\,nm and $W_i$ = 3\,$\mu$m one obtains $\delta \approx 2.6$ and $\lambda_{J\rm{,i}} \approx 50\,\mu$m. Thus, $l \approx$ 1.6, justifying the short junction approach taken above. An ambiguity, however, arises from the problem to refer the measured $I_c$ to the proper junction area. Above, we have used the whole ramp area. Alternatively, assuming that the current is dominantly carried by the in-plane currents on the YBCO side one might refer to a junction area which is projected perpendicular to the substrate plane. Then, $j_c$ increases by a factor of 7 and $\lambda_J$ decreases by a factor of 3, bringing the junction closer to the long junction regime. To distinguish these scenarios we simulated $I_c(H)$ based on the sine-Gordon equations for the case of $\theta^* = 90^\circ$ and various values of the normalized junction length $l$. For the flux density profile the same shape as in the inset of Fig.~\ref{fig:int_gen} (a) has been used. Fig. \ref{fig:short-vs-long} shows simulations of $I_c(H)$ patterns for different $l$ and the calculation using the short junction model. 
Deviations from the short junction model that are incompatible with our experimental data  become prominent near the main $I_c$ maxima for $l>3$ (the case of $l$ = 4 is shown in the graph). By contrast, $I_c(H)$ for $l$ = 1.6 is almost indistinguishable from the short junction model and in agreement 
with the above estimate, using the ramp area instead of its projection.
\par


\subsection{Conclusions}
In summary, we have investigated the interference patterns $I_c$ vs. $H$ for a 8 facet YBCO-Au-Nb zigzag-ramp Josephson junction, with a facet length of 10\,$\mu$m. The angle $\theta$ between the substrate plane and the applied field $H$ was systematically varied and we also discussed two in-plane angles $\phi$ between $H$ and the facet orientation ($\phi=0^\circ$ and $\phi=45^\circ$). All interference patterns could be understood from the fact that a field component perpendicular to the substrate causes a strong and spatially varying contribution to the flux density profile inside the junction. The junction by itself -- admittedly our best junction - seems to be very homogeneous, with an essentially constant critical current density inside each facet. Particularly, no $j_c$ asymmetry between facets oriented along the in-plane $x$ and $y$ directions were observable. 
We have expected a strong dephasing effect on $I_c$ vs. $H$ due to the fact that the flux penetrating the junction is not preserved. The effect is present but at least an order of magnitude weaker than expected.  Our investigations also showed that $H$ is applied ``parallel'' to the junction (in terms of an homogeneous flux density profile) for an angle $\theta_h\lesssim 2^\circ$ 
and not for a field orientation about parallel to the ramp angle $\theta_r = 8^\circ$. These findings may contribute new knowledge to the general physics of ramp junctions.

Further, due to strong compression of the perpendicular field component by about a factor of 100, only for angles $|\theta^*| \equiv |\theta-\theta_h|\ll0.1^\circ$ the interference patterns were ``ideal'' in the sense that the flux density in the junction is essentially homogeneous. The effective junction thickness is only about 85\,nm, leading to enormous fields ($\sim$ 35\,mT) that are required to produce a flux quantum per (\opi)\,-\,segment for the case of $\theta^* = 0^\circ$. We also demonstrated, that there is a dead angle $\theta^*_d$ very close to ``parallel'' field orientation, where tremendous changes in $I_c(H)$ occur. These properties make it extremely hard to study zigzag junctions in ``parallel'' field configuration. Due to idle regions the Josephson length of our junction was on the order of 50\,$\mu$m, requiring junction lengths of several 100\,$\mu$m to study long junction effects. For such junctions the field compression will be even bigger than in our case, and, thus, the field alignment must be much better than $0.1^\circ$ relative to $\theta^*=0$ to achieve a homogeneous flux density. Thus it seems that, for a study of (semi)fluxon physics\cite{Hilgenkamp03,Susanto05,Cedergren:2010:TTGB:0-pi-JJ:Semifluxon,Cedergren:2010:TTGB:0-pi-JJ} or $\varphi$-junction effects \cite{Buzdin03,Goldobin07a,Zazunov09,Goldobin11} in zigzag junctions magnetic fields should be oriented perpendicular to the substrate plane and the corresponding flux density profiles should be taken into account rather than being avoided by parallel field alignment.

We finally note that dead angles, as described here also appear in other types of Josephson junctions or Josephson junction arrays\cite{Heinsohn01,Monaco09,Scharinger10}. Whenever there is a strong compression of the perpendicular field component this angle may be very close to the parallel field orientation, making measurements in ``parallel'' field unreliable, favoring the perpendicular field orientation.

\par
\acknowledgments
We acknowledge financial support by the Deutsche Forschungsgemeinschaft (Project KO 1303/10) and by the German Israeli Foundation (Grant No. G-967-126.14/2007).
\par
\bibliography{SF,SIFS_REFs}
\end{document}